\documentclass[apj,numberedappendix]{emulateapj}
\usepackage{amsmath}

\def\gtorder{\mathrel{\raise.3ex\hbox{$>$}\mkern-14mu
             \lower0.6ex\hbox{$\sim$}}}
\def\ltorder{\mathrel{\raise.3ex\hbox{$<$}\mkern-14mu
             \lower0.6ex\hbox{$\sim$}}}

\slugcomment{Accepted to ApJ}

\shorttitle{A mixture evolution scenario of the AGN RLF}

\shortauthors{Yuan et al.}

\begin{document}

\title{A mixture evolution scenario of the AGN radio luminosity function}

\author{
Zunli Yuan\altaffilmark{1,2}, Jiancheng Wang\altaffilmark{1,2}, Ming Zhou\altaffilmark{1,2}, Jirong Mao\altaffilmark{1,2}
}
\email{yuanzunli@ynao.ac.cn}

\altaffiltext{1} {Yunnan Observatories, Chinese Academy
of Sciences,  Kunming 650011, China}
\altaffiltext{2} {Key Laboratory for the Structure and Evolution of Celestial Objects,
Chinese Academy of Sciences,  Kunming 650011, China}

\begin{abstract}

We propose a mixture evolution scenario to model the evolution of the radio luminosity function (RLF) of steep spectrum AGNs (active galactic nuclei)  based on a Bayesian method. In this scenario, the shape of the RLF is determined by both the density and luminosity evolution. Our models indicate that the density evolution is positive until a redshift of $\thicksim 0.9$, at which point it becomes negative, while the luminosity evolution is positive to a higher redshift ($z \thicksim 5$ for model B and $z \thicksim 3.5$ for model C), where it becomes negative. Our mixture evolution model works well, and the modeled RLFs are in good agreement with previous determinations. The mixture evolution scenario can naturally explain the luminosity-dependent evolution of the RLFs.

\end{abstract}


\section{Introduction}

Observations have suggested that the radio-loud active galactic nuclei (AGNs) can play an important role in feedback, and thus have a significant impact on galaxy evolution, with AGN outflows being responsible for controlling or terminating star formation
\citep[SF; e.g.][]{2009Natur.460..213C,2012ARA&A..50..455F,2014MNRAS.445..955B,2014ARA&A..52..589H}. To understand the significance of such AGN feedback processes and their influence on the global SF history of the Universe \citep{2013MNRAS.436.1084M}, it is important to first understand the evolution of the radio luminosity function (RLF) to high redshift.

In the study of RLF, an increasingly acceptable viewpoint is that the powerful radio AGNs undergo very rapid evolution to a redshift of $z\sim 3$ , while the lower luminosity radio population only experiences much milder positive evolution to $z\sim 1$ \citep[e.g.][]{2001MNRAS.328..882W,2004MNRAS.352..909C,2011MNRAS.416.1900R,2013MNRAS.436.1084M, 2015A&A...581A..96R}. The different evolution for high- and low-luminosity radio-loud AGNs may reminds one to regard them as two essentially different populations. However, the simple division based on the radio luminosity may not reveal the physical essence. Over recent years, it has become clear that a more essential classification of radio-loud AGNs is based on the AGN accretion state. Radio sources powered by radiatively efficient accretion of cold gas onto a geometrically thin, optically thick accretion disk are referred to as high-excitation radio galaxies (HERGs), or `quasar-mode' accretors \citep[][]{2012MNRAS.421.1569B,2013MNRAS.430.3086G}. While those powered by the radiatively inefficient accretion of hot gas \citep{2007MNRAS.376.1849H,2014ARA&A..52..529Y} onto a geometrically thick accretion disk are referred to as low-excitation radio galaxies (LERGs), or `radio-model' accretors. Quasar-mode and radio-model AGNs dominate the radio-AGN population at higher and lower radio luminosities respectively \citep{2014MNRAS.445..955B}. However, this does not mean one can separate the two radio populations by a simple division in radio luminosity. As shown by \citet{2012MNRAS.421.1569B}, both quasar-mode and radio-mode radio AGNs are found across all radio luminosities.

In the past two decades, research on the RLF of radio-loud AGNs has been increasingly abundant\citep[e.g.][]{2001MNRAS.328..882W,2004MNRAS.352..909C,2009ApJ...696...24S,2009MNRAS.392..617D,2013MNRAS.436.1084M,2015A&A...581A..96R,
2015MNRAS.452.2692F}. However, most of these works are non-parametric RLFs derived by the classical binned $1/V_{max}$ method\citep[see][]{1968ApJ...151..393S,2013Ap&SS.345..305Y}. The two most representative papers \citep{1990MNRAS.247...19D,2001MNRAS.322..536W} on the parametric analysis of RLFs were published more than a decade ago. The above authors artificially divided the RLFs to be the sum of two components, i.e., a high-luminosity evolving component $\rho_h$, and a low-luminosity non-evolving/mild-evolving component $\rho_l$. Then they considered $\rho=\rho_l+\rho_h$ and modeled $\rho$ using pure density evolution (PDE) or pure luminosity evolution (PLE) models. In this work, we do not artificially separate the low-luminosity and high-luminosity populations into two evolving components. We use mixture evolution models, which consider a combination of density and luminosity evolution, to parameterize the steep-spectrum AGN RLFs. For the parameterizing procedure, a Bayesian method is used.

Throughout the paper, we adopt a Lambda Cold Dark Matter cosmology with the parameters $\Omega_{m}$ = 0.27,  $\Omega_{\Lambda}$ = 0.73, and $H_{0}$ = 71 km s$^{-1}$ Mpc$^{-1}$.

\section{The sample}
The data we used in this work is the combined sample established by \citet{2012ApJ...744...84Y}. It consists of four sub-samples: the MRC1 \citep{1996ApJS..107...19M}, the MS4 \citep{2006AJ....131..100B}, the BRL \citep{1999MNRAS.310..223B} and the 3CRR \citep{1983MNRAS.204..151L} samples. The sources (totaling 1063) in our sample are all steep-spectrum ones, which mainly consist of radio galaxies (RGs) and steep-spectrum quasars. All of the sources have flux density data observed at 408 MHz. At this relatively lower frequency, the radio emission is mainly from the extended lobes of the radio-loud AGNs, minimizing the effect of orientation bias. For more details about the combined sample, please refer to \citet{2012ApJ...744...84Y}.

\section{Methods}

\subsection{The Likelihood function for the RLF}

The RLF $\rho(z,L)$ is defined as the number of sources per comoving volume $V(z)$ with radio luminosities in the range of $L,L+dL$. It is related to the probability distribution of $(z,L)$ by
\begin{eqnarray}
\label{eqpc}
p(z,L)=\frac{1}{N}\rho(z,L)\frac{dV}{dz}.
\end{eqnarray}
where $N$ is the total number of sources in the universe, and is given by the integral of $\rho$ over $L$ and $V(z)$ \citep{2008ApJ...682..874K}. The likelihood function for the observed data $p(L_{obs},z_{obs}|\theta)$can be derived, once we assume a parametric form for $\rho(z,L)$, with parameters $\theta$. \citet{1983ApJ...269...35M} give a likelihood function based on the Poisson distribution and define $S=-2ln(p(L_{obs},z_{obs}|\theta))$. Dropping terms independent of the model parameters, one finds
\begin{eqnarray}
\label{likelihood1}
\begin{aligned}
S=-2\sum_{i}^{N_{obs}}&ln[\rho(z_{i},L_{i})]+\\
&2\int\int\rho(z,L)\Omega(z,L)\frac{dV}{dz}dzdL.
\end{aligned}
\end{eqnarray}
Usually, best estimates for the model parameters are obtained by minimizing $S$. In this work, we use a Bayesian method as described in section \ref{Bayes} to obtain not only the best estimates for the model parameters, but also their probability distribution.
The limits of the integral in $S$ should consider our combined sample, which consists of four sub-samples. Then we have
\begin{eqnarray}
\label{likelihood2}
\begin{aligned}
S=-2\sum_{i}^{N_{obs}}&ln[\rho(z_{i},L_{i})]+\\
&\sum_{j}^{4}\Omega^j\int^{z_2^j}_{z_1^j}dz\frac{dV}{dz}\int_{max[L_1^j,L_{min}^j(z)]}^{L_2^j}\rho(z,L)dL,
\end{aligned}
\end{eqnarray}
where $(z_1^j,z_2^j)$ and $(L_1^j,L_2^j)$ are the redshift and luminosity limits of the $j$th sub-sample, respectively, $L_{min}^j(z)$ is the luminosity limit corresponding to the flux density limit, and $\Omega^j$ is the solid angle subtended by the sub-sample j.

Thus, we have given the general form of the RLF likelihood function. It can be used to perform Bayesian inference by combining with a prior distribution.

\begin{figure*}[!htb]
\centering
\includegraphics[width=1.02\columnwidth]{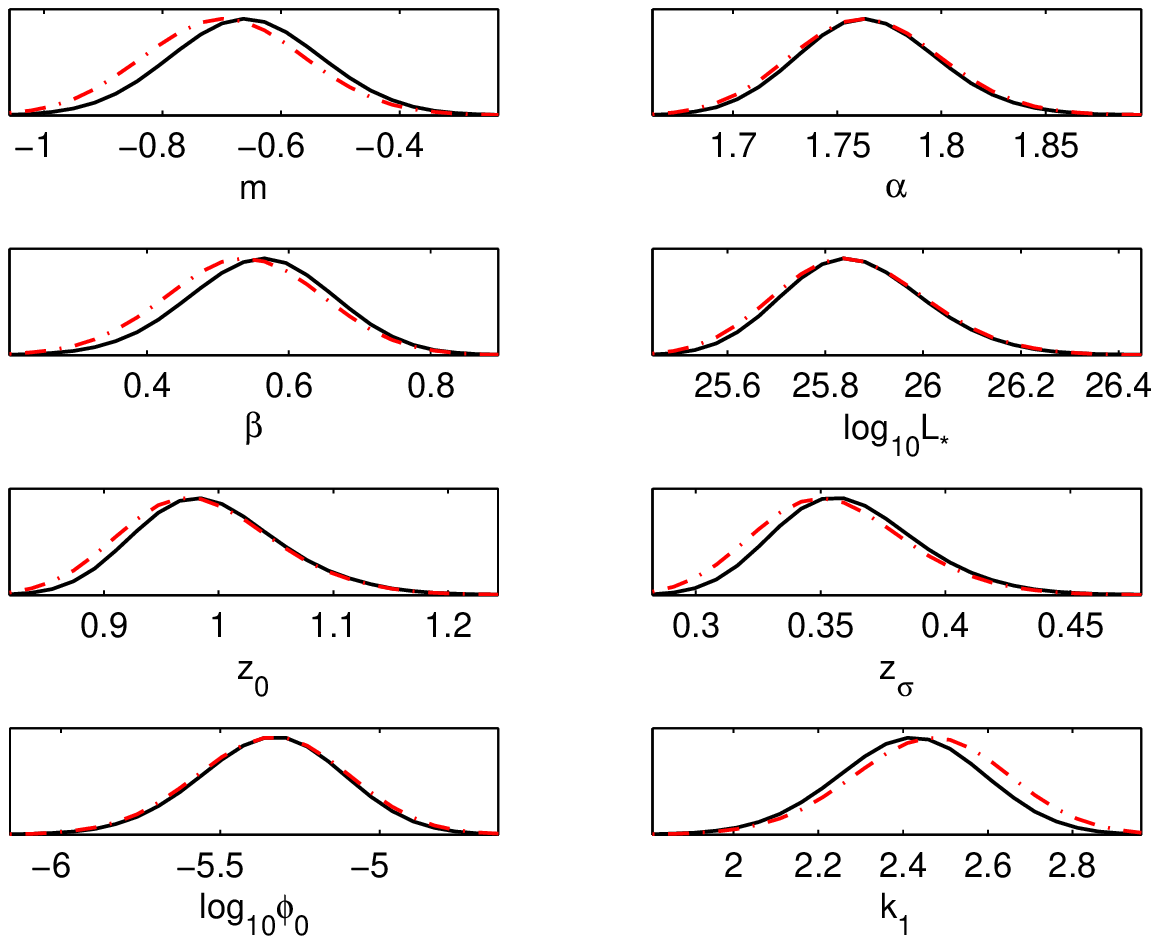}
\includegraphics[width=1.02\columnwidth]{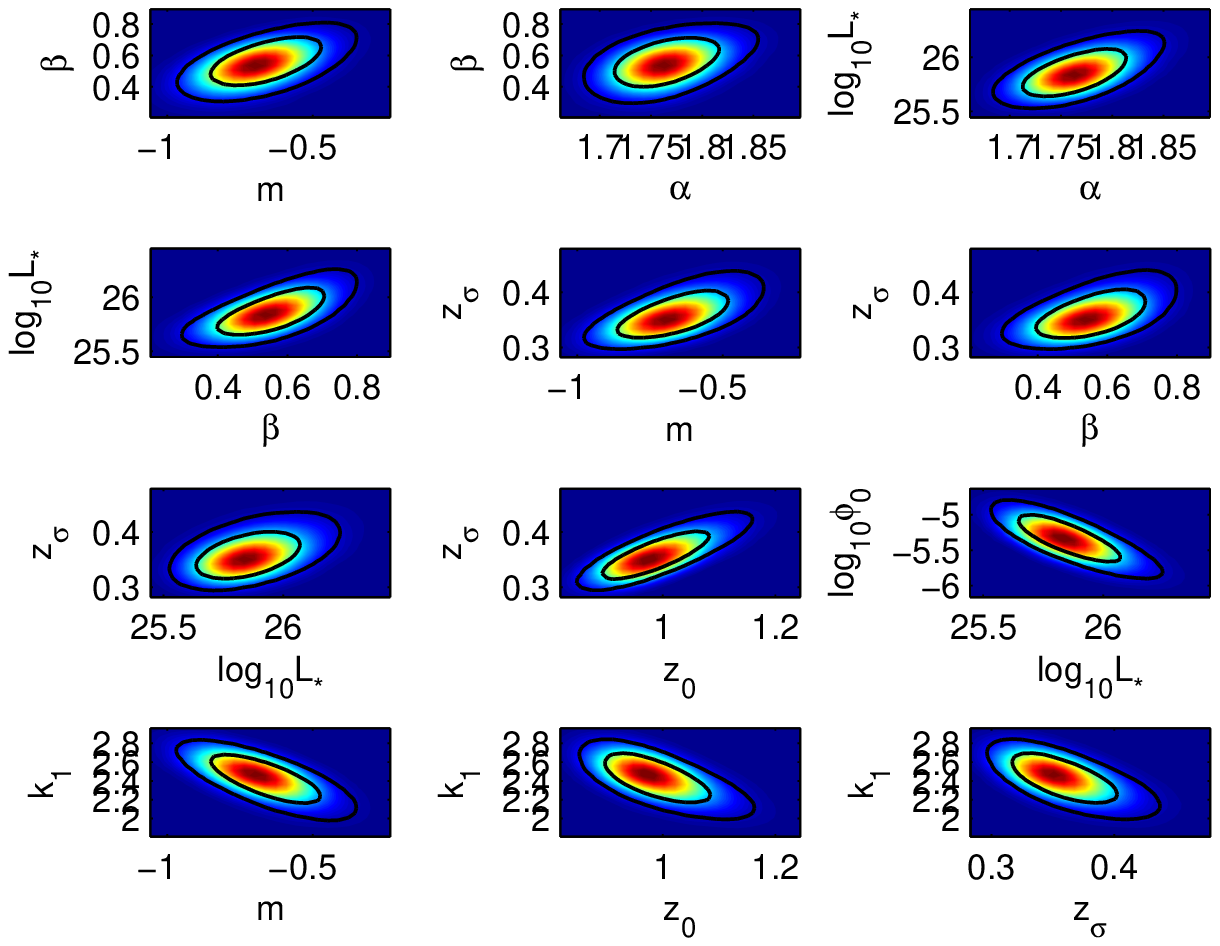}
\includegraphics[width=1.02\columnwidth]{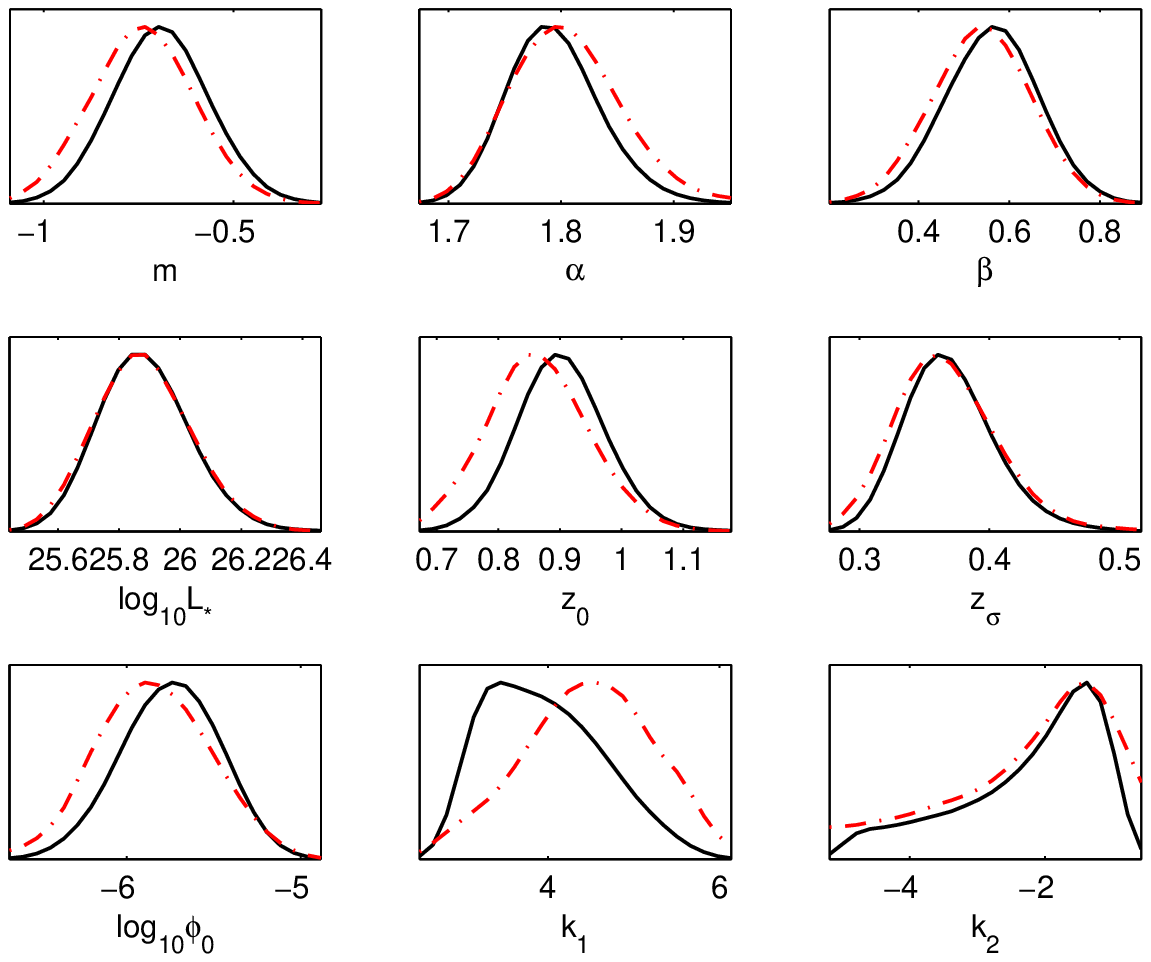}
\includegraphics[width=1.02\columnwidth]{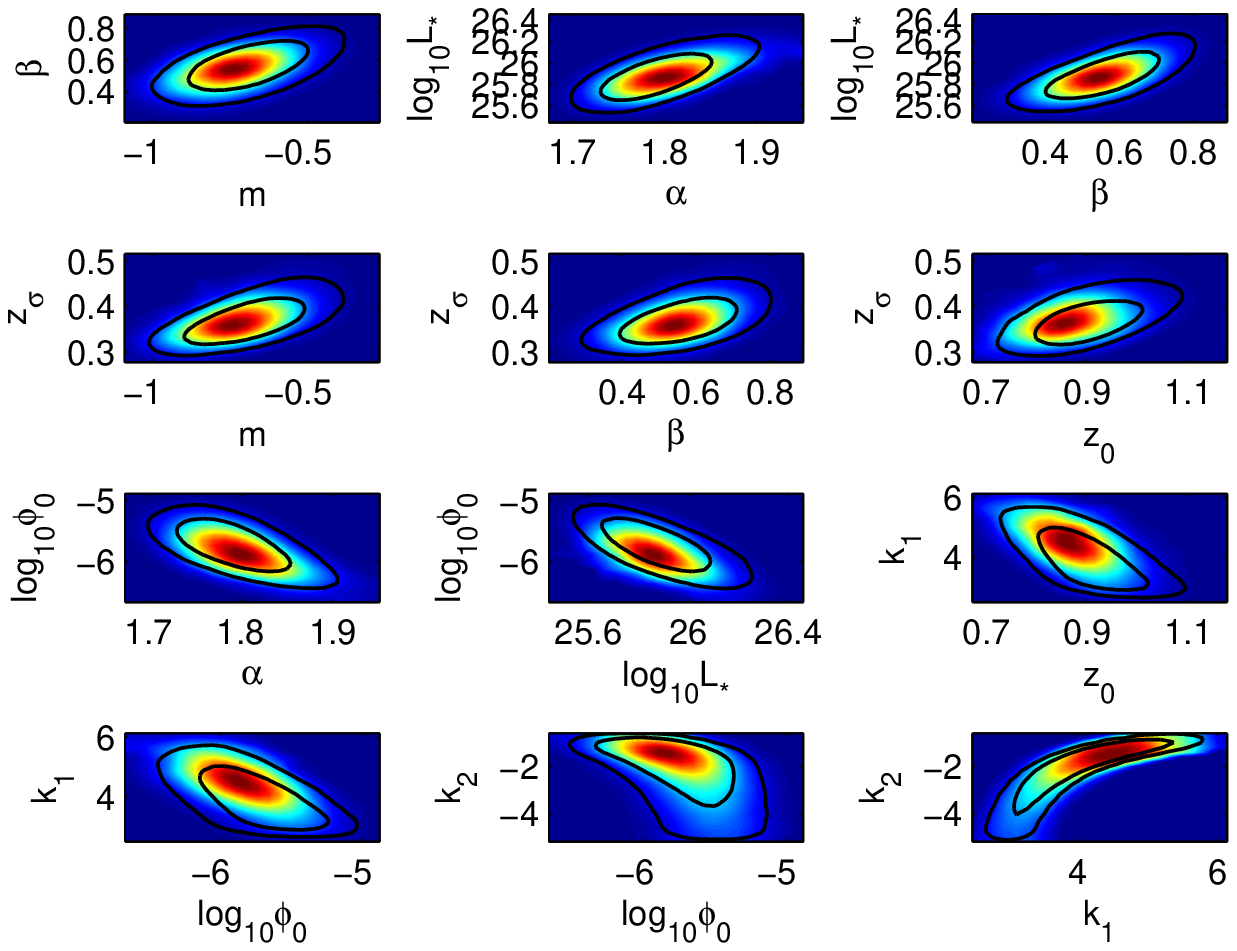}
\includegraphics[width=1.02\columnwidth]{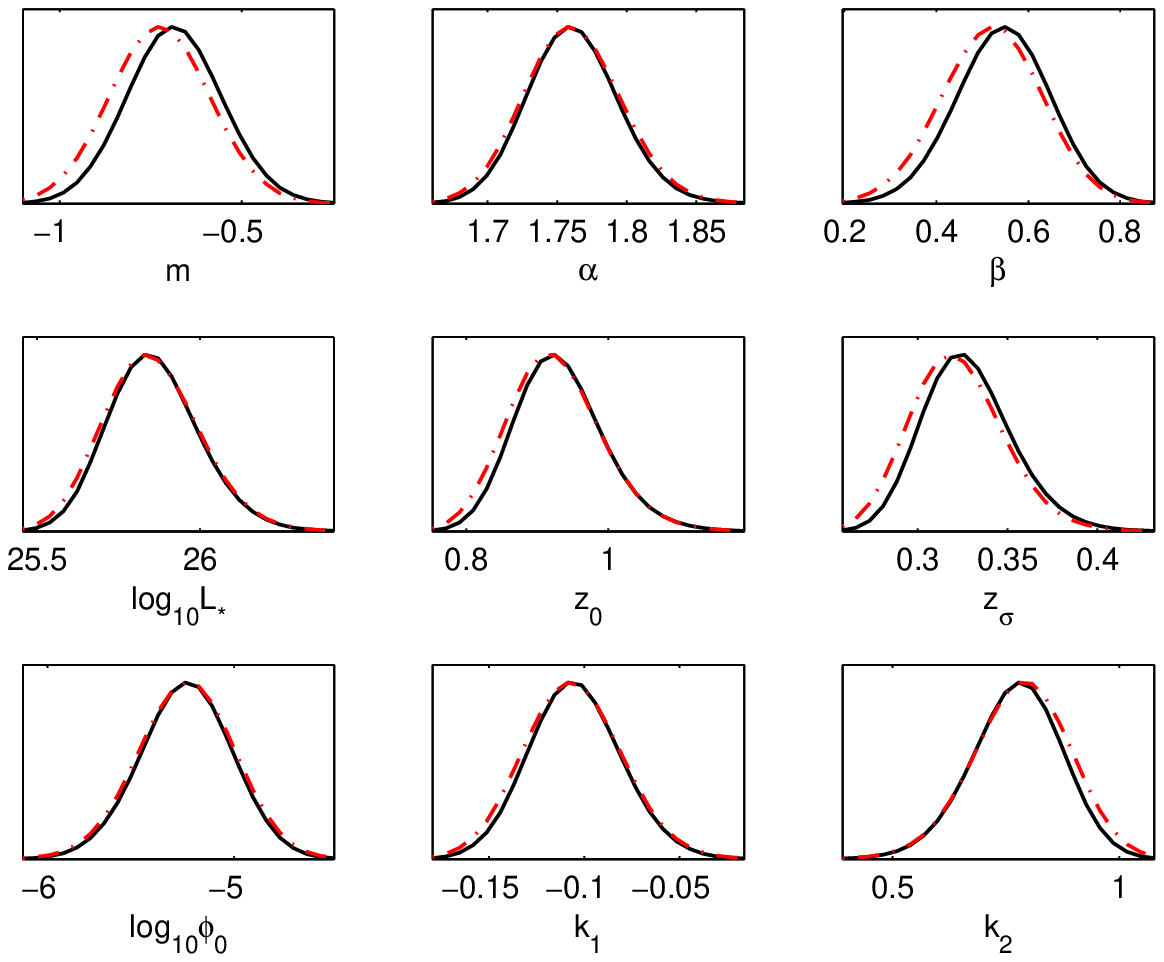}
\includegraphics[width=1.02\columnwidth]{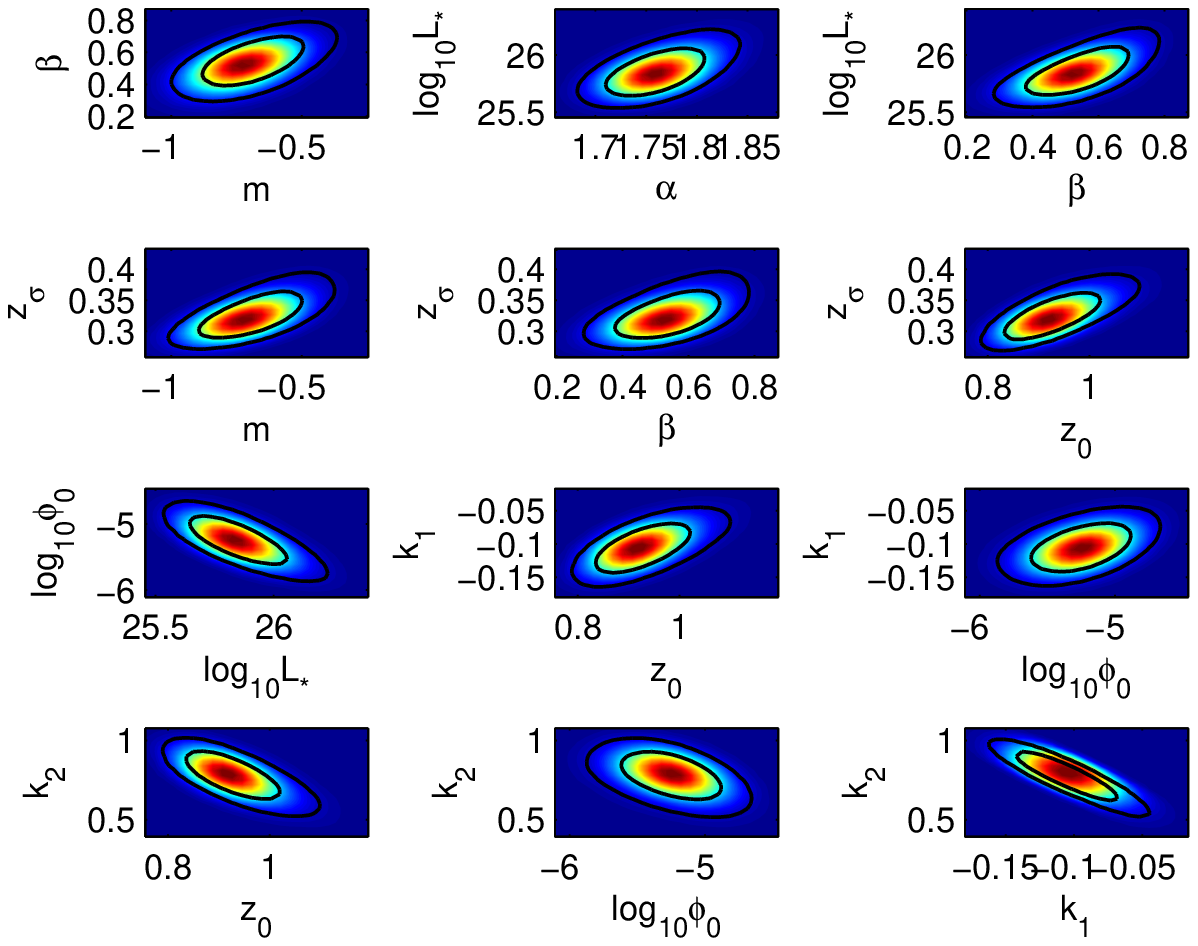}
\caption{Left: 1-D probability distribution of the parameters in the models \citep[the red dash dot curves are the mean likelihoods of MCMC samples and the black solid curves are the marginalized probabilities, see][]{2002PhRvD..66j3511L}; Right: 2-D confidence contours of the parameters. The contours are for 1 and 2 $\sigma$ levels. For the 2D confidence regions of the parameters, we only show combinations with relatively large correlation. The upper, middle and lower panels are for model A, C and D respectively.}
\label{fig:mcmc}
\end{figure*}

\begin{figure}
  \centerline{
    \includegraphics[scale=0.40,angle=0]{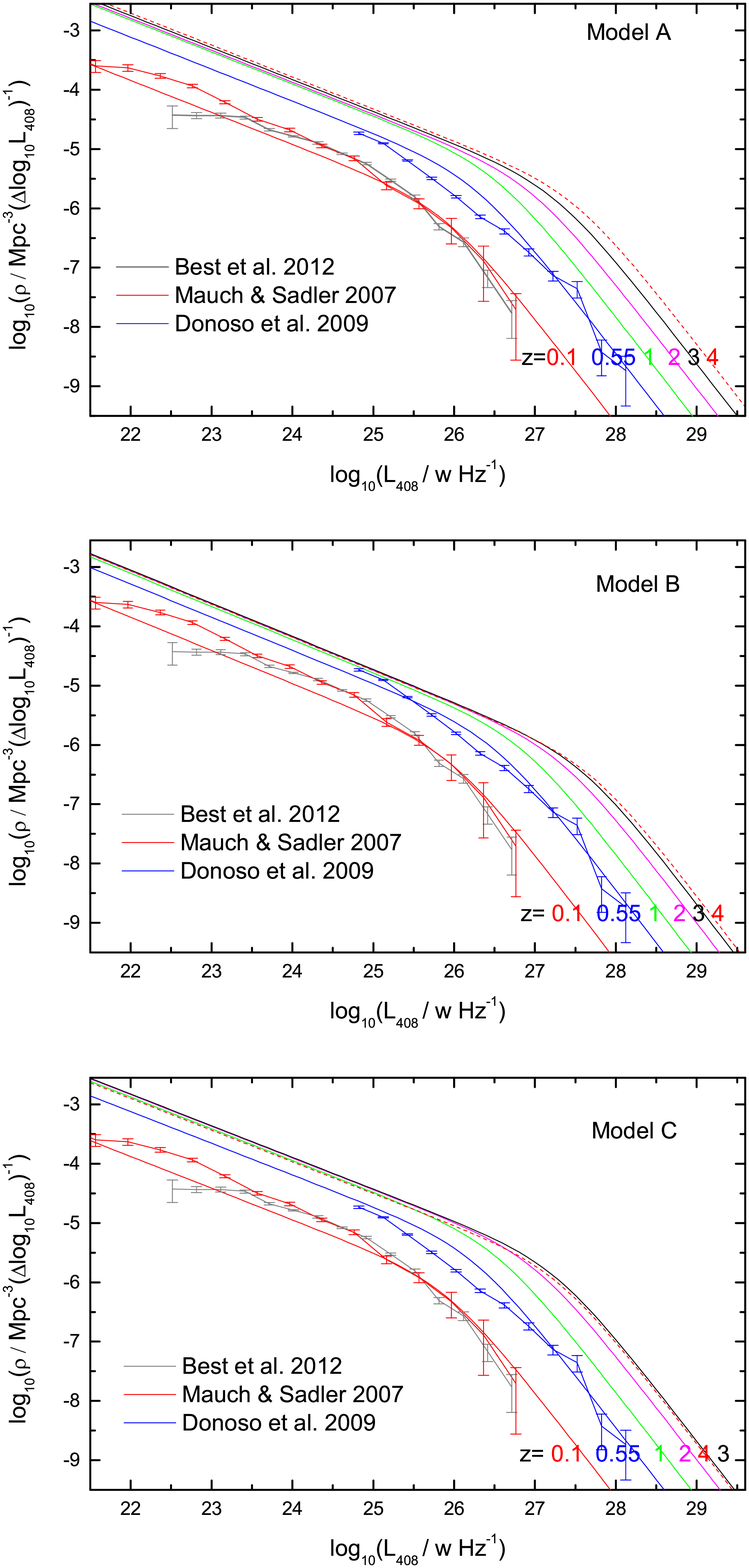}}
  \caption[R-z relation]{\label{f3} The RLFs derived for model A (top panel), B (middle panel) and C (bottom panel) at z=0.1,0.55,1,2,3 and 4 (red, blue, green, violet, black solid lines and red dashed lines respectively). The RLFs at z=4 are marked as red dashed lines to clarify whether a redshift cut-off occurs or not at z$>$3. The red and gray lines with error bars show the local RLFs of \citet{2012MNRAS.421.1569B} and \citet{2007MNRAS.375..931M} respectively. The blue lines with error bars show the RLF at $\langle z \rangle$ =0.55 derived by \citet{2009MNRAS.392..617D}. These RLFs have been converted to the same frequency and cosmology used by our work.}
\label{RLF}
\end{figure}

\subsection{Bayesian inference for the RLF parameters}
\label{Bayes}

The purpose of Bayesian analysis is to estimate the probability distribution of the model parameters (called as the posterior distribution), given by the observed data. It is
\begin{eqnarray}
\begin{aligned}
p(\theta|L_{obs},z_{obs})\varpropto p(\theta)p(L_{obs},z_{obs}|\theta),\\
&
\end{aligned}
\end{eqnarray}
where $p(\theta)$ is the prior distribution of $(\theta)$, which should convey information known prior to the analysis, and $p(L_{t,obs},z_{obs},I|\theta)$ is the observed data likelihood function defined in Equations \ref{likelihood1} and \ref{likelihood2}.
In this work, we assume a uniform prior on $\theta$. This is equivalent to setting a value range for it.

The posterior distribution of $\theta$ given above can be used to perform Bayesian inference. This can be achieved by drawing values of $\theta$ from its posterior distribution using a Markov chain Monte Carlo (MCMC) sampling algorithm. The MCMC sampler here used is the public code ``COSMOMC", which was presented by \citet{2002PhRvD..66j3511L}. See \citet{2003itil.book.....M} for more details about the MCMC method.

\subsection{The form of the RLF}

For the forms of the RLFs, pure density evolution (PDE) models \citep[e.g.,][]{2001MNRAS.322..536W} and pure luminosity evolution (PLE) models \citep[][]{1990MNRAS.247...19D} have been adopted in the literature. To achieve a good fitting, the above authors have to consider the RLFs to be the sum of two components, i.e., a high-power evolving component $\rho_h$, and a low-power non-evolving/mild-evolving component $\rho_l$. We use the mixture evolution models first introduced by \citet{1988ApJ...325..103H} to model the RLF. The mixture evolution models consider a combination of density and luminosity evolution to parameterize the RLF. Mathematically, suppose the evolution of RLF is a vector $\vec{E}$, then it can be written as

\begin{eqnarray}
\label{vector}
\vec{E}=e_1\vec{E}_d+e_2\vec{E}_l,
\end{eqnarray}
where $\vec{E}_d$ and $\vec{E}_l$ are the base vectors of density and luminosity evolution respectively. $e_1$ and $e_2$ are the coefficients, which are usually the functions of redshift. The physical meaning of density evolution is whether the sources are more/less numerous than that of today, while the luminosity evolution represents whether the sources are more/less luminous than that of today. Thus the advantage of mixture evolution models is that they can discriminate the above two evolutions with different physical meaning.

Our mixture evolution model RLF can be written as the following general form.
\begin{eqnarray}
\label{eqn:rhoform}
\rho(z,L_{t})=e_1(z)\rho(z=0,L/e_2(z)),
\end{eqnarray}
where $e_1(z)$ and $e_2(z)$ are the functions describing the density and the luminosity evolution respectively.
The density evolution function $e_1(z)$ we used is very similar to the one adopted by \citet{2008ApJ...674..111C},
\begin{eqnarray}\label{eqn:LFPLEnoev}
e_1(z) =
\begin{cases}
  \displaystyle z^m\exp\left[-\frac{1}{2}\left(\frac{z-z_0}{z_{\sigma}}\right)^2\right],  &  0<z \leqslant z_0 \\
  z^m, & z>z_0
\end{cases}
\end{eqnarray}
where $m$, $z_0$, and $z_{\sigma}$ are free parameters of the model.
The local luminosity function $\rho(z=0,L/e_2(z=0))$ is a double power-law form described as
\begin{eqnarray}
\label{eqn:rhot}
\begin{aligned}
\rho(z=0,L/e_2(z&=0))=\frac{dN}{d\log L} \\
&=\phi_0\left[\left(\frac{L}{L_*}\right)^{\alpha} + \left(\frac{L}{L_*}\right)^{\beta}\right]^{-1},
\end{aligned}
\end{eqnarray}
where $\phi_0$, $\alpha$, $L_*$ and $\beta$ are free parameters of the model. The luminosity evolution function $e_2(z)$ has three different forms depending upon the model: the traditional power-law form
\begin{eqnarray}
e_2(z)=(1+z)^{k_1}
\end{eqnarray}
for model A, the modified power-law form used by \citet{2012ApJ...751..108A}
\begin{eqnarray}
e_2(z)=(1+z)^{k_1}\mathrm{exp}(z/k_2)
\end{eqnarray}
for model B, and the polynomial form used by \citet{1990MNRAS.247...19D} and \citet{2000MNRAS.317.1014B}
\begin{eqnarray}
\label{eqn:flc}
e_2(z)=10^{k_1z+k_2z^2}
\end{eqnarray}
for model C.

\section{results}

\subsection[]{The modeled RLFs}
In Figure \ref{fig:mcmc}, we show the one-dimensional (1D) probability distributions \citep[the dashed curves are the mean likelihoods of MCMC samples and the solid curves are the marginalized probabilities, see][]{2002PhRvD..66j3511L,2015MNRAS.454.1310Y} and two-dimensional (2D) confidence regions (at 1$\sigma$ and 2$\sigma$ confidence levels) of the model parameters. The upper, middle, and lower panels are for models A, B, and C respectively. The best-fit model parameters correspond to the peak of the 1D probability distributions (the marginalized probabilities, the solid curves in Figure \ref{fig:mcmc}). For the 2D confidence regions of the parameters, we only show the combinations with relatively large correlations \citep[e.g.,][]{2011ApJ...735..120Y}. The values of the best-fitting parameters and their 1 $\sigma$ error are reported in Table \ref{tab:fit}.

\begin{table*}
\tablewidth{0pt}
\renewcommand{\arraystretch}{1.8}
\caption[]{Best-fit Parameters}
\begin{center}
\begin{tabular}{lccccccccc}
\hline\hline

\colhead{Model}           & \colhead{$\log_{10}\phi_0$} &
\colhead{$\log_{10}L_*$}  & \colhead{$\alpha$} &
\colhead{$\beta$}         & \colhead{m} &
\colhead{z$_{\rm 0}$}     & \colhead{z$_{\rm \sigma}$} &
\colhead{k$_1$}           & \colhead{k$_2$} \\
\hline

A & -5.278$_{-0.280}^{+0.151}$ & 25.82$_{-0.09}^{+0.18}$ & 1.76$_{-0.03}^{+0.03}$ & 0.54$_{-0.07}^{+ 0.12}$ & -0.67$_{-0.11}^{+0.13}$ & 0.974$_{-0.041}^{+0.080}$ & 0.351$_{-0.016}^{+0.038}$ & 2.434$_{-0.185}^{+0.148}$ & - \\

B & $-5.836_{-0.191}^{+0.367}$ & $25.87_{-0.13}^{+0.15}$ & $1.79_{-0.03}^{+0.04}$ & $0.56_{-0.09}^{+0.10}$ & $-0.72_{-0.09}^{+0.15}$  & $0.876_{-0.041}^{+0.093}$ & $0.365_{-0.027}^{+0.037}$ & $4.321_{-1.068}^{+0.361}$ & $-1.496_{-1.953}^{+0.259}$\\

C & $-5.308_{-0.182}^{+0.286}$ & $25.86_{-0.13}^{+0.13}$ & $1.76_{-0.03}^{+0.03}$ & $0.53_{-0.08}^{+ 0.11}$ & $-0.73_{-0.07}^{+0.17}$  & $0.915_{-0.041}^{+0.076}$ & $0.320_{-0.015}^{+0.032}$ & $-0.107_{-0.021}^{+0.025}$ & $0.796_{-0.119}^{+0.072}$\\
\hline
\end{tabular}
\end{center}
~~Units -- $\phi_0$: [${\rm Mpc^{-3}}$],\,\, $L_*$: [${\rm W Hz^{-1}}$]. The best-fitting parameters as well as their 1 $\sigma$ errors for model A, B and C.
\label{tab:fit}
\end{table*}

\begin{table}
\renewcommand{\arraystretch}{1.0}
\caption[]{Goodness-of-fit}
\begin{center}
\begin{tabular}{lccc}
\hline\hline

\colhead{Model} & \colhead{1D-$P_{\mathrm{KS}}$-$L$} & \colhead{1D-$P_{\mathrm{KS}}$-$z$}  &  \colhead{2D-$P_{\mathrm{KS}}$}   \\
\hline
 A &0.58 & 0.99 & 0.34 \\
 B &0.67 & 0.99 & 0.39 \\
 C &0.56 & 0.99 & 0.33 \\
\hline
\end{tabular}
\end{center}
\label{tab:ks}
The goodness-of-fit of the models. These are the probabilities from the 1D (for $z$ and $L$) and 2D KS tests to the $L_{408}-z$ distribution.
\end{table}

In Figure \ref{RLF}, we show the best-fit RLFs of models A, B, and C at $z=0.1, 0.55, 1, 2, 3$ and $4$. All three models show luminosity-dependent evolution where the low-luminosity radio sources experience weaker evolution than their bright counterparts. The faint sources experience mild positive evolution out to $z \thicksim$ 1.0. Beyond this redshift, this evolution slows down. However, the bright sources evolve strongly to $z > 3.0$. The above pictures are in good agreement with previous studies, showing that our mixture evolution models are effective.

Figure \ref{RLF} shows that our three models present very similar results for the bright end RLFs, while they give slightly different results for the faint end RLFs. Model B gives a milder evolution for the faint sources than that given by Models A and C. We then perform the 1D and 2D Kolmogorov-Smirnov (KS) tests \citep{1983MNRAS.202..615P,1987MNRAS.225..155F,1992nrfa.book.....P} to estimate the goodness-of-fit of models. The goodness-of-fit (KS test probabilities) of the models are reported in Table \ref{tab:ks}. All the three models are acceptable because they satisfy the condition of $P_{\mathrm{KS}}\geqslant0.2$ \citep{1983MNRAS.202..615P}. The KS tests show that model B is marginally favored over models A and C.

\subsection[]{Comparing with local RLFs}
Up to now, multi-wavelength observed data of the radio-loud AGNs has become abundant. By combining different surveys, such as the Sloan Digital Sky Survey (SDSS), the NRAO (National Radio Astronomy Observatory) VLA (Very Large Array) Sky Survey (NVSS) and the Faint Images of the Radio Sky at Twenty Centimetres (FIRST) survey, one is able to construct large samples of radio-loud AGNs down to a flux limit of few mJy. Based on these large and deep samples, the local \citep{2007MNRAS.375..931M,2012MNRAS.421.1569B} and low ($z\thicksim0.55$) redshift \citep{2009MNRAS.392..617D} RLFs are well determined. Figure \ref{RLF} shows the comparison among these results with our model RLFs. Notably, all of these previous RLFs are calculated at 1.4 GHz. We have shifted luminosities from 1.4 GHz to 408 MHz according to a power law ($L/\nu \propto \nu^{-\alpha}$, with $\alpha=0.70$). Besides, we also need to convert their results to the one in the same cosmology used by our work. This can be achieved by using the following relation from \citet{1985MNRAS.217..601P},

\begin{eqnarray}
\label{convert}
\rho_1(z,L_1)\frac{dV_1}{dz}=\rho_2(z,L_2)\frac{dV_2}{dz}.
\end{eqnarray}

Figure \ref{RLF} shows both the local and $z=0.55$ RLFs are in good agreement with previous determinations, indicating that our mixture evolution model is successful.

\begin{figure}[!htb]
\centering
\includegraphics[width=1.02\columnwidth]{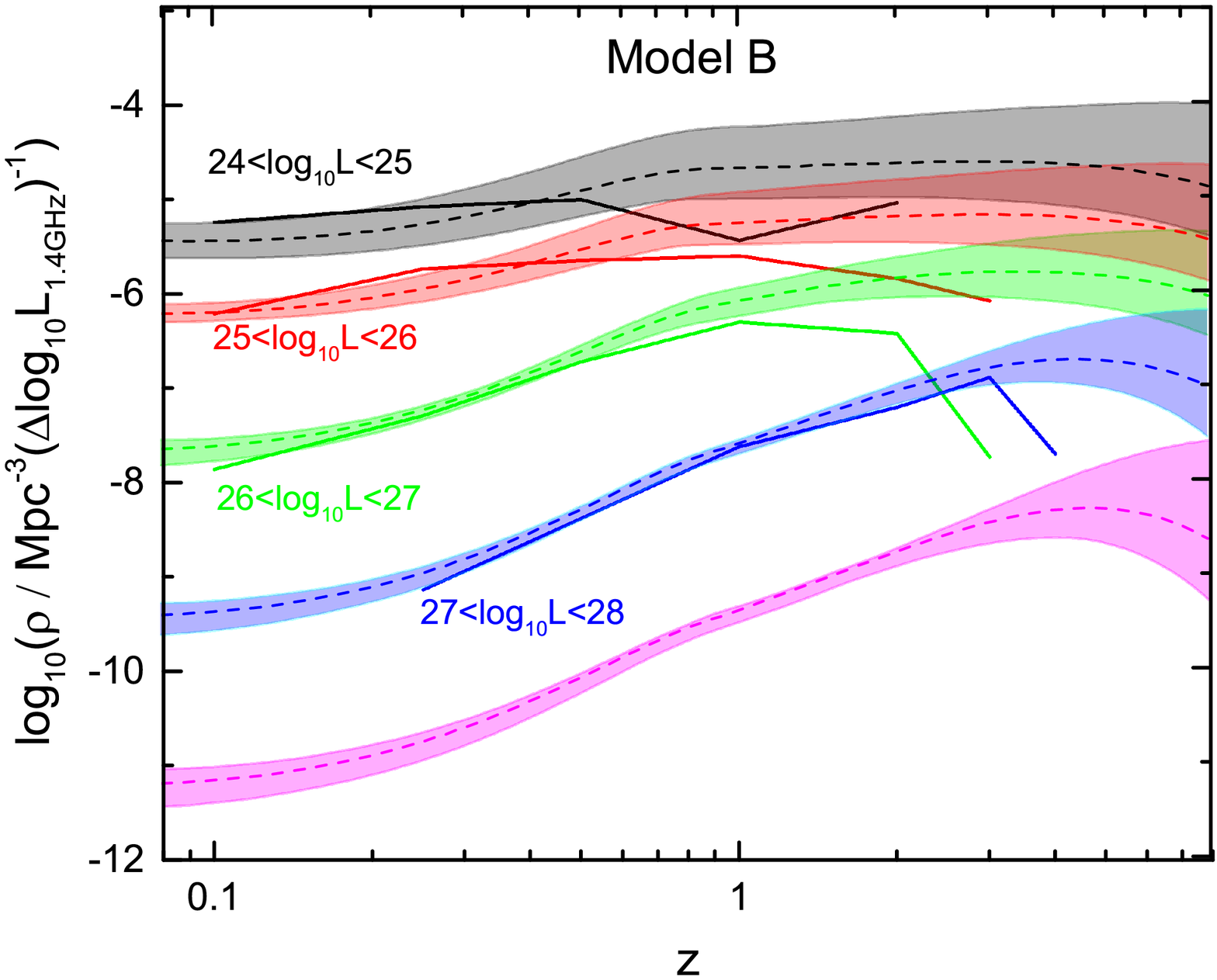}
\includegraphics[width=1.02\columnwidth]{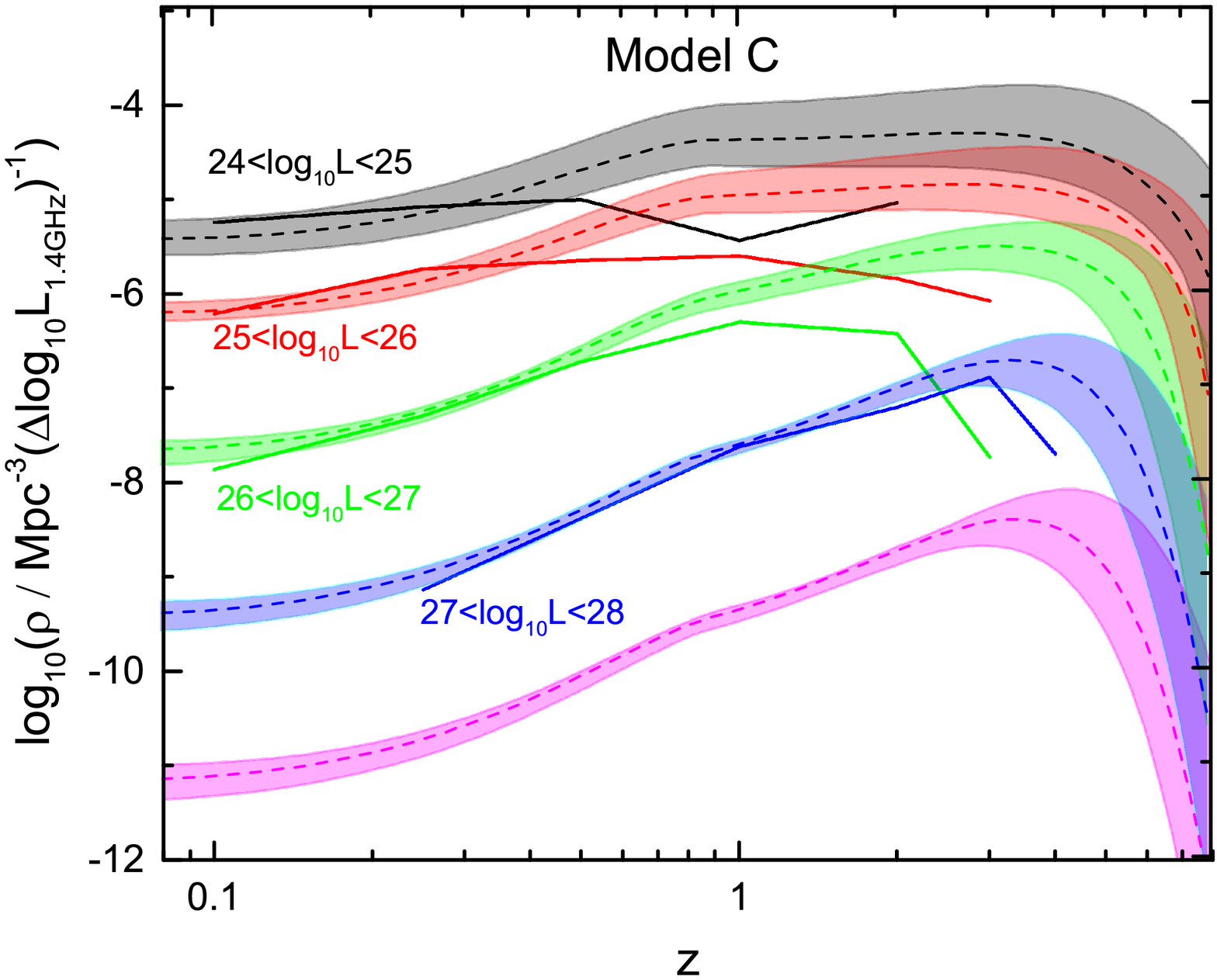}
\caption{Model B and Model C RLFs as function of redshift, having been converted to the same frequency used by \citet{2015A&A...581A..96R}. The black, red, green, blue and violet dashed lines show the RLFs at $\log_{10}L_{1.4\mathrm{GHz}}$=24.5, 25.5, 26.5, 27.5 and 28.5 respectively. The light shaded areas take into account the 1 $\sigma$ error bands. The black, red, green and blue solid lines show the result of \citet{2015A&A...581A..96R}.}
\label{luminosity-dependence}
\end{figure}

\begin{figure}
  \centerline{
    \includegraphics[scale=0.45,angle=0]{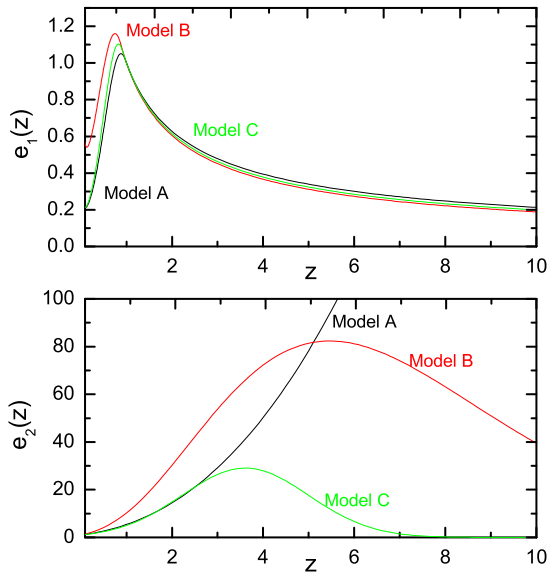}}
  \caption[A example LF]{\label{f2} Density and luminosity evolution functions for Models A (black solid lines), B (red solid lines) and C (green solid lines).}
\label{e1e2}
\end{figure}

\subsection[]{Luminosity-dependence of the high-redshift turnover}
In recent years, evidence is growing that the position of the steep-spectrum RLF peak is luminosity dependent, being interpreted as a sign of cosmic downsizing. Recent evidence is given by \citet{2015A&A...581A..96R}, showing that the space density of the most powerful sources peaks at higher redshift than that of their weaker counterparts. Figure \ref{luminosity-dependence} shows the  space density changing with redshift for our model B and C at various luminosities. For the convenience in comparing with the result of  \citet{2015A&A...581A..96R} (their Figure 4), we have shifted the luminosities to 1.4 GHz by assuming a spectral index of 0.70. It shows that our models are in broad agreement with their result. We also notice that a space density peak is more clearly shown at high radio luminosities than that at low radio luminosities. Similar phenomenon is also noticed by \citet{2015A&A...581A..96R}, who find that at $P_{1.4 \mathrm{GHz}} > 10^{26}$ W/Hz the redshift of the peak space density increases with luminosity, whilst at lower radio luminosities the position of the peak remains constant within the uncertainties.

\subsection[]{The Redshift Cut-off}

The redshift cut-off, involving whether the comoving number density of radio AGNs declines dramatically at the redshift beyond z $\sim$ 2.5, is a unsolved problem in the research of RLF. The RLFs of flat-spectrum radio sources support the existence of the redshift cut-off \citep[e.g., ][]{1990MNRAS.247...19D,1996Natur.384..439S,2000MNRAS.319..121J,2005A&A...434..133W}, while the redshift cut-off in the steep-spectrum radio sources has much controversy. \citet{1990MNRAS.247...19D} first proposed a red-shift cutoff for steep-spectrum radio sources, but several authors \citep[e.g., ][]{2001MNRAS.327..907J,2007MNRAS.375.1349C} subsequently argued that there is no compelling evidence of a high-redshift decline in steep spectrum radio sources.

Considering the existence of peaks for density and luminosity evolution (see Figure \ref{e1e2}), both model B and C in principle enable the occurrence of redshift cut-off. However, the evidence for a redshift cut-off does not appear to be significant given the uncertainties on the space density evolution shown in Figure \ref{luminosity-dependence} for Model B. In comparison, Model C support a more significant evidence for a redshift cut-off. Nevertheless, we can notice that the turnover redshift is at least greater than 4, which is remarkably bigger than previous determinations \citep[$z \sim$ 2.5, e.g., ][]{1990MNRAS.247...19D}. To sum up, we believe that the existence of redshift cut-off still depends on models. Given that the only model (C) which supports a significant existence of redshift cut-off is not overwhelmingly favoured over the others, we conclude that based on the current sample we do not find compelling evidence for a redshift cut-off for steep-spectrum radio sources. In the future, samples based on deeper and fainter radio surveys are expected to provide more significant constraint on the density evolution, especially the luminosity evolution of RLFs.

\begin{figure}
  \centerline{
    \includegraphics[scale=0.40,angle=0]{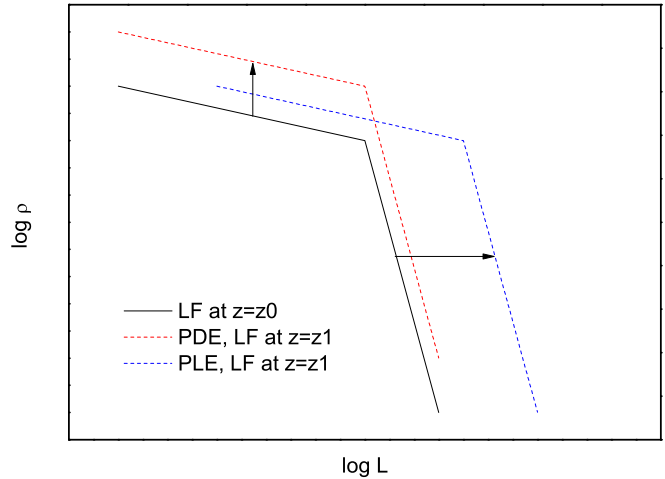}}
  \caption[A example LF]{\label{f2} Example of LF shows that the flatter an LF is, the more sensitive is it to the density evolution, whereas the steeper an LF is, the more sensitive is it to the luminosity evolution.}
\label{example_RLF}
\end{figure}

\section[]{Discussion}

The main point of this work is that we use mixture evolution to model the RLFs. In this scenario, the shape of RLFs is determined by both the density and luminosity evolution. Our result (see Figure \ref{e1e2}) indicates that the positive density evolution continues to
a redshift of $\thicksim 0.9$ and then changes its sign (positive to negative), while the positive luminosity evolution continues
to a higher redshift ($\thicksim 5$ for model B and $\thicksim 3.5$ for model C) and then changes its sign. Usually, the flatter a LF is,
the more sensitive is it to the density evolution, whereas the steeper a LF is, the more sensitive is it to the luminosity evolution (see Figure \ref{example_RLF}). For a double power-law RLF, it is generally flatter at low luminosity and steeper at high luminosity, hence the turnover redshift of evolution for low-luminosity sources must be lower than that of high-luminosity sources,
displaying a luminosity-dependent evolution.

Intuitively understanding, the physical meaning of density evolution is whether the sources are more/less numerous than that of today, while the luminosity evolution represents whether the sources are more/less luminous than that of today. But the more profound physical significance behind them is still unknown. We speculate that the density evolution is associated with the density distribution of accreting black holes, while the luminosity evolution is related with the changing of accretion state. This will be the subject of a future work.

\section[]{Conclusions}

The main results of this work are as follows.

\begin{enumerate}
  \item We propose a mixture evolution scenario to model the steep-spectrum AGN RLF based on a Bayesian method. In this scenario, the shape of the RLF is determined together by the density and luminosity evolution. Our models indicate that the positive density evolution continues to a redshift of $\thicksim 0.9$ and then changes its sign (positive to negative), while the positive luminosity evolution continues to a higher redshift ($\thicksim 5$ for model B and $\thicksim 3.5$ for model C) and then changes its sign.
  \item The mixture evolution scenario can naturally explain the luminosity-dependent evolution of the RLF. In agreement with previous results, we find that a space density peak is more obviously shown at high radio luminosities than that at low radio luminosities.
  \item The existence of redshift cut-off still depends on models. Given that the only model (C) that supports a significant existence of redshift cut-off is not overwhelmingly favored over the others, we conclude that based on the current sample we do not find compelling evidence for a redshift cut-off for steep-spectrum radio sources. Even if the redshift cut-off indeed exists, we believe the turnover redshift should be greater than previous determinations.
\end{enumerate}

\section*{Acknowledgments}
We are grateful to the referee for useful comments that improved this paper. We acknowledge the financial supports from the National Natural Science Foundation of China 11133006, 11163006, 11173054,11573060, the Policy Research Program of Chinese Academy of Sciences (KJCX2-YW-T24) and CAS "Light of West China" Program. J.M. is supported by the Hundred-Talent Program of Chinese Academy of Sciences and the Key Research Program of Chinese Academy of Sciences (grant No. KJZD-EW-M06). The authors gratefully acknowledge the computing time granted by the Yunnan Observatories, and provided on the facilities at the Yunnan Observatories Supercomputing Platform.

\end{document}